\documentclass[twocolumn,prl,showpacs,preprintnumbers]{revtex4-1}%
\usepackage{amssymb}
\usepackage{amsmath}
\usepackage{graphicx}
\usepackage{epsfig}
\usepackage{subfigure}
\usepackage{amsfonts}
\usepackage{CJK}
\usepackage{appendix}
\usepackage{multirow}
\usepackage{amsfonts}
\usepackage{dsfont}
\usepackage{amsmath}
\usepackage{graphicx}
\usepackage{float}
\usepackage{amsmath,epsfig,amssymb,subfigure,bm,dsfont}
\usepackage{lipsum}
\usepackage[english]{babel}
\usepackage{bm}
\usepackage{amsmath}
\usepackage[colorlinks,linkcolor=blue,citecolor=blue]{hyperref}%
\providecommand{\U}[1]{\protect \rule{.1in}{.1in}}
\begin{document}

\title{Boundary-induced Phases in the Dissipative Dicke Lattice Model}
\author{Peng-Fei Wei$^{1}$}
\author{Yilun Xu$^{2,3}$}
\author{Fengxiao Sun$^{2,3}$}
\author{Qiongyi He$^{2,3,4}$}
\author{Peter Rabl$^{5,6,7}$}
\author{Zhihai Wang$^{1}$}
\email{wangzh761@nenu.edu.cn}
\affiliation{1. Center for Quantum Sciences and School of Physics, Northeast Normal University, Changchun 130024, China\\
2. State Key Laboratory for Mesoscopic Physics, School of Physics,
Frontiers Science Center for Nano-optoelectronics, Peking University, Beijing 100871, China\\
3. Beijing Academy of Quantum Information Sciences, Beijing 100193, China\\
4. Collaborative Innovation Center of Extreme Optics, Shanxi University, Taiyuan 030006, China\\
5. Walther-Mei\ss ner-Institut, Bayerische Akademie der Wissenschaften, 85748 Garching, Germany\\
6. Technische Universit\"at M\"unchen, TUM School of Natural Sciences, Physics Department, 85748 Garching, Germany\\
7. Munich Center for Quantum Science and Technology (MCQST), 80799 Munich, Germany}

\begin{abstract}
The superradiant phase transition in the dissipative Dicke lattice model, driven by on-site collective atom-photon interactions and inter-site photon hopping, is a cornerstone of nonequilibrium quantum many-body physics. However, little is still known about the influence of boundaries in experimental achievable systems of finite size. Here we investigate the dissipative superradiant phase transition in the Dicke lattice model with a small number of sites and reveal a striking sensitivity of this model to the nature of the boundary conditions. Specifically, we find that under open boundary conditions a whole zoo of superradiant phases with broken translational symmetry appears, which is not observed in the corresponding infinite lattice system.  Our results demonstrate the crucial influence of boundary effects on the stationary phases of dissipative lattice models, which offers intriguing new opportunities for studying these phenomena in near term experimental realizations of such models in quantum optics and circuit QED.
\end{abstract}

\maketitle

{\bf Introduction}-Quantum phase transitions (QPTs) refer to qualitative changes in the physical properties of a system when a control parameter approaches a critical value~\cite{XB1997,XB1999,XB2004,XB2006,XB2007,XB2009,XB2020,XB2023}. While traditionally studied in condensed matter systems, the concept of QPTs has been successfully extended to the field of quantum optics. A paradigmatic example is provided by the Rabi and Dicke models~\cite{Dicke1954,Dicke19731,Dicke19732,Dicke2003PRL,Dicke2003PRE,Dicke20102,Dicke2014,Rabi2015,Rabi2017}, where a single or multiple two-level systems are strongly coupled to a single-mode resonator field, which is observed in various physical platforms experimentally~\cite{DickeEX2010,DickeEX2014,Rabi2021,DickeEX2015,DickeEX2021,DickeEX2023}. Since no quantum system is entirely isolated from its environment, the study of dissipative QPTs in the open Dicke model has attracted significant attention~\cite{Dicke2010,Dicke2017,Dicke2018,Dicke2019,Rabi2023,Dicke2023}. In particular, both the ground state of the closed system and the steady state of the open system exhibit a transition from the normal phase to a superradiant phase as the light-matter coupling strength crosses a critical value.

Meanwhile, the single-site Dicke model has been extended to multi-site systems~\cite{ND2014,JCL2016,Rabi2021PRL,Rabi2022PRL,threeD2022,twoD2024,twoD2025,threeD2025}, forming the so-called Dicke lattice model, where inter-site photon hopping significantly modifies the critical on-site atom-photon coupling strength. Although any physical realization of a photonic lattice is necessarily finite in size, the impact of boundary conditions---specifically, periodic boundary conditions (PBC) versus open boundary conditions (OBC)---on the dissipative phase transition in the Dicke lattice model has received surprisingly little theoretical attention. In particular, a systematic understanding of how boundary-induced asymmetries influence the phase structure and critical behavior in finite-size quantum open systems remains lacking, leaving a critical gap in the current theoretical framework.

To address this gap, we investigate the dissipative superradiant phase transition in a Dicke lattice model. Specifically, we present the full phase diagram for a lattice with $N=3$ sites and demonstrate that OBC leads to a more intricate structure compared to PBC. Under OBC, the homogeneous superradiant phase is completely absent, the critical atom-photon coupling strength is modified, and the range of photonic hopping strengths that ensure dynamical stability is significantly shifted. These exotic features induced by OBC persist in lattices with larger, yet still finite, numbers of sites.

\begin{figure}
    \centering
    \includegraphics[width=\linewidth]{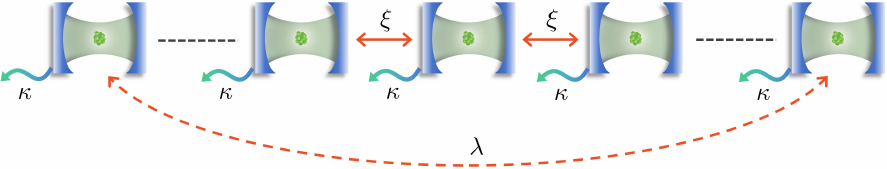}
    \caption{Schematic of the open Dicke lattice model. The system consists of an array of lossy resonators, each coupled to an ensemble of two-level atoms. The parameter $\xi$ denotes the photon hopping strength between nearest-neighbor resonators. The boundary condition is characterized by $\lambda = \xi$ for periodic boundary conditions and $\lambda = 0$ for open boundary conditions.
}
    \label{fig:modeltu}
\end{figure}

{\bf Dicke lattice model}-As schematically illustrated in Fig.~\ref{fig:modeltu}, we consider a waveguide consisting of $N$ resonators, each of which interacts with an ensemble of $N_a$ identical two-level atoms. This system is described by the Dicke lattice model with Hamiltonian
\begin{eqnarray}
H &=& \sum_{i=1}^{N} H_{i}^{\rm Dicke} - \sum_{i=1}^{N-1} \xi \left( c_{i}^{\dagger} c_{i+1} + c_{i} c_{i+1}^{\dagger} \right) \nonumber \\&&- \lambda \left( c_{1}^{\dagger} c_{N} + c_{1} c_{N}^{\dagger} \right),
\end{eqnarray}
where the local Dicke Hamiltonian for the $i$th lattice site is given by
\begin{eqnarray}
H_{i}^{\rm Dicke} = \omega_c c_i^{\dagger} c_i + \omega_{a} S_i^z + \frac{2g}{\sqrt{N_a}} ( c_i + c_i^\dagger ) S_i^x.
\end{eqnarray}
Here, $c_i$ is the annihilation operator for the $i$th resonator with frequency $\omega_c$, and $S_i^{x,y,z} = \sum_{j=1}^{N_a} \sigma_{i,j}^{x,y,z}/2$ are the collective spin operators for the atomic ensemble in the $i$th lattice site, with transition frequency $\omega_a$. The parameter $g$ denotes the on-site atom-photon coupling strength, while $\xi$ and $\lambda$ represent the inter-site photon hopping strengths between neighboring resonators and between the first and last resonators, respectively.

In the following, we consider both PBC and OBC, corresponding to $\lambda = \xi$ and $\lambda = 0$, respectively. Under these conditions, the dispersion relations of the photonic modes are derived in the Supplementary Materials (SM)~\cite{SM}, yielding
\begin{align}
\omega_{P,k} &= \omega_c - 2\xi \cos\left[ \frac{2\pi(k-1)}{N} \right], \\
\omega_{O,k} &= \omega_c - 2\xi \cos\left[ \frac{\pi k}{N+1} \right],
\end{align}
with $k = 1, 2, \dots, N$. To ensure dynamical stability, the frequencies must remain positive, i.e., $\omega_{P(O),k} > 0$, which permits the hopping strength $\xi$ to take either positive or negative values. In the main text, we restrict our analysis to the regime $\xi > 0$, while the case of $\xi < 0$ is discussed in the SM~\cite{SM}.

To account for photon losses into free space, we describe the system dynamics using the Lindblad master equation
\begin{equation}
\frac{d\rho}{dt} = -i[H, \rho] + \kappa \sum_{i=1}^{N} \left( 2 c_i \rho c_i^{\dagger} - c_i^{\dagger} c_i \rho - \rho c_i^{\dagger} c_i \right),
\end{equation}
where $\kappa$ denotes the photonic decay rate. For simplicity, spontaneous emission from the atoms is neglected.

Applying the mean-field approximation in the thermodynamic limit ($N_a \rightarrow \infty$), the dynamical equations for the expectation values of the relevant operators are given by
\begin{equation}
\begin{aligned}
\frac{d}{dt}\langle c_{j} \rangle &= -(\kappa + i\omega_c)\langle c_{j} \rangle - i\frac{g}{\sqrt{N_a}} ( \langle S_{j}^{+} \rangle + \langle S_{j}^{-} \rangle ) \\
&\quad + i\xi \sum_{i=1}^{N-1} ( \delta_{i,j} \langle c_{i+1} \rangle + \delta_{i+1,j} \langle c_i \rangle ) \\
&\quad + i\lambda ( \delta_{1,j} \langle c_N \rangle + \delta_{N,j} \langle c_1 \rangle ), \\
\frac{d}{dt}\langle S_{j}^{-} \rangle &= -i\omega_a \langle S_{j}^{-} \rangle + i\frac{2g}{\sqrt{N_a}} ( \langle c_{j}^{\dagger} \rangle + \langle c_j \rangle ) \langle S_{j}^{z} \rangle, \\
\frac{d}{dt}\langle S_{j}^{z} \rangle &= -i\frac{g}{\sqrt{N_a}} ( \langle c_{j}^{\dagger} \rangle + \langle c_j \rangle ) ( \langle S_{j}^{+} \rangle - \langle S_{j}^{-} \rangle ).
\end{aligned}
\label{eqs:wentai}
\end{equation}

Using the spin conservation relation $|\langle S_{j}^{-} \rangle|^2 + \langle S_{j}^{z} \rangle^2 = N_a^2 / 4$, one can numerically obtain the steady-state solutions by setting $d\langle \mathcal{O} \rangle / dt = 0$. The stability of these solutions can be further analyzed using the Routh-Hurwitz criterion~\cite{QZZL1987}, as detailed in SM~\cite{SM}. Consequently, the steady-state phase diagram in the $\xi$-$g$ parameter plane can be constructed.

In the SM~\cite{SM}, we also analytically derive the critical on-site atom-photon coupling strengths $g_c^{P,NP}$ and $g_c^{O,NP}$, which separate the superradiant phases (SRP) from the normal phase (NP) under PBC ($\lambda = \xi$) and OBC ($\lambda = 0$), respectively. These are given by
\begin{equation}
g_c^{P(O),NP} = \min_k \frac{1}{2} \sqrt{ \omega_a \omega_{P(O),k} \left( 1 + \frac{\kappa^2}{\omega_{P(O),k}^2} \right) }.
\label{phaseboundary}
\end{equation}

For small coupling strength $g < g_c^{P(O),NP}$, the system settles into the NP, characterized by $\langle c_j \rangle = \langle S_j^{-} \rangle = 0$ and $\langle S_j^z \rangle = -N_a/2$ for all $j = 1, 2, \dots, N$. In contrast, for $g > g_c^{P(O),NP}$, at least one of expectation values $\langle c_j \rangle$ becomes non-zero, spontaneously breaking the $\mathbb{Z}_2$ symmetry of the master equation. This regime corresponds to the SRP.

\begin{figure}
    \centering
    \includegraphics[width=\linewidth]{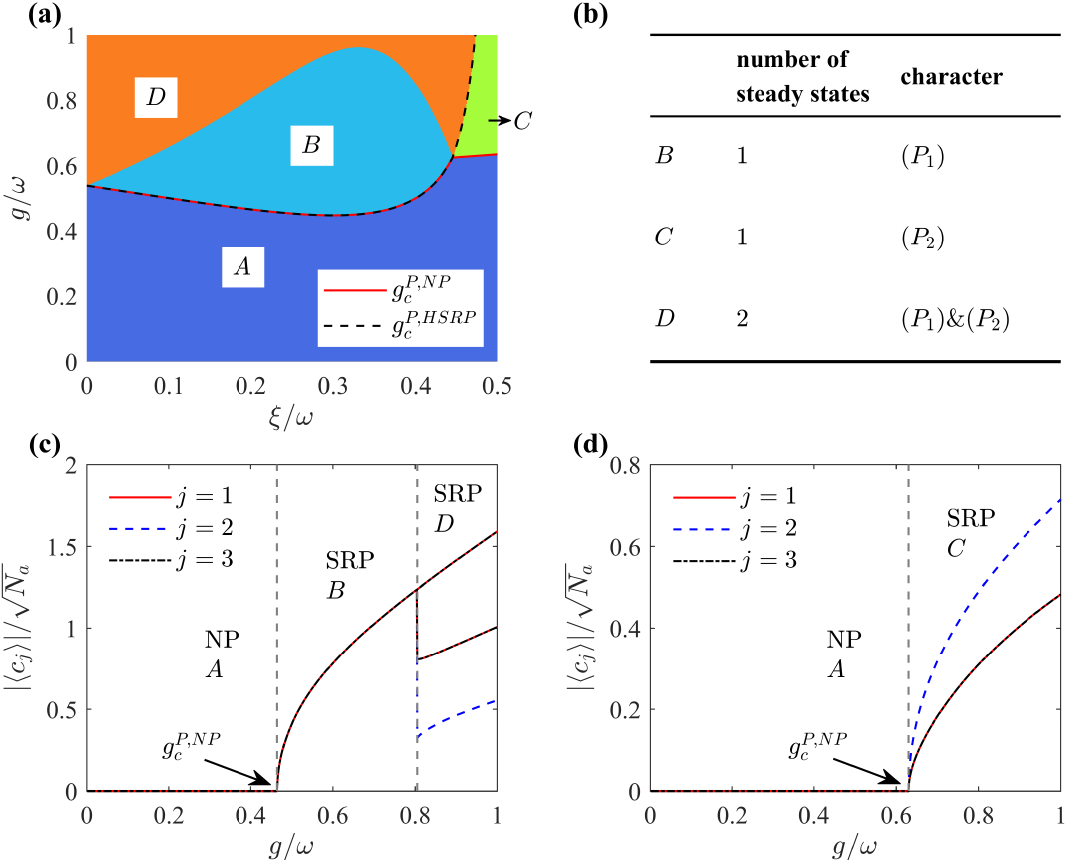}
    \caption{(a) Phase diagram and (b) its characterization for the Dicke lattice model with \( N = 3 \) under PBC. (c,d) Plot of the order parameter \( |\langle c_j \rangle|/\sqrt{N_a}\,(j=1,2,3) \) as a function of the coupling strength \( g/\omega \), with (c) \( \xi = 0.2\omega \) and (d) \( \xi = 0.48\omega \). Other parameters are set to \( \omega_a = \omega_c = \omega = 1 \) and \( \kappa = 0.4\omega \).}
    \label{fig:xiangtuPBC}
\end{figure}

\begin{figure}
    \centering
    \includegraphics[width=\linewidth]{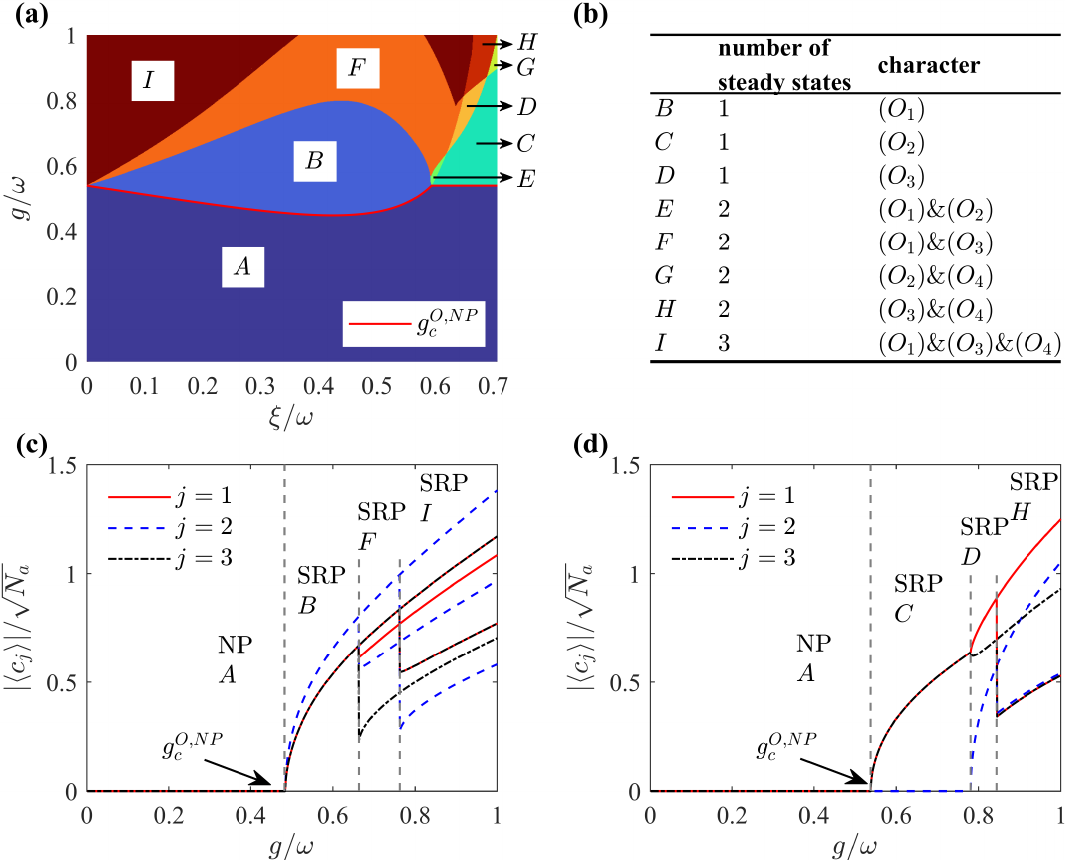}
    \caption{(a) Phase diagram and (b) its characterization for the Dicke lattice model with \( N = 3 \) under OBC. (c,d) Plot of the order parameter \( |\langle c_j \rangle|/\sqrt{N_a}\,(j=1,2,3) \) as a function of the coupling strength \( g/\omega \), with (c) \( \xi = 0.2\omega \) and (d) \( \xi = 0.67\omega \). Other parameters are the same as in Fig. \ref{fig:xiangtuPBC}.}
    \label{fig:xiangtuOBC}
\end{figure}

{\bf Phase diagram for $N=3$}-For the minimal model with $N=3$, where the boundary condition can be clearly defined, we numerically obtain the steady-state phase diagrams shown in Fig.~\ref{fig:xiangtuPBC}(a) and Fig.~\ref{fig:xiangtuOBC}(a), corresponding to the PBC and OBC, respectively. In both cases, we can find the NP $A$, and the critical coupling strengths, $g_c^{P,NP}$ and $g_c^{O,NP}$,  they separate the NP and SRPs and are indicated by the red solid curves. The detailed character of the SRPs are illustrated in Fig.~\ref{fig:xiangtuPBC}(b) and Fig.~\ref{fig:xiangtuOBC}(b).

Under PBC, as shown in Fig.~\ref{fig:xiangtuPBC}(a) and (b), we identify three distinct SRPs, denoted by $B$, $C$, and $D$. The phase $B$ corresponds to a homogeneous SRP, where the steady-state expectation values are given by~\cite{SM}
\begin{equation}
\begin{aligned}
\frac{\mathcal{C}}{\sqrt{N_a}} &= \pm \frac{\omega_a}{4g} \left(1 + \frac{i\kappa}{\omega_{P,1}}\right) \sqrt{ \frac{N_a^2}{4\mathcal{Z}^2} - 1 }, \\
\mathcal{S} &= \pm \mathcal{Z} \sqrt{ \frac{N_a^2}{4\mathcal{Z}^2} - 1 }, \\
\frac{\mathcal{Z}}{N_a} &= -\frac{\omega_a \omega_{P,1}}{8g^2} \left( 1 + \frac{\kappa^2}{\omega_{P,1}^2} \right).
\end{aligned}
\label{hom}
\end{equation}
In this phase, translational symmetry is preserved, as evidenced by the relation
\((\mathcal{C}, \mathcal{S}, \mathcal{Z}) := (\langle c_1 \rangle, \langle S_1^- \rangle, \langle S_1^z \rangle) = (\langle c_2 \rangle, \langle S_2^- \rangle, \langle S_2^z \rangle) = (\langle c_3 \rangle, \langle S_3^- \rangle, \langle S_3^z \rangle\)),
indicating that all three sites are identical. This configuration is labeled as $P_1$ in the first row of Fig.~\ref{fig:xiangtuPBC}(b).

In contrast, phase $C$ corresponds to an inhomogeneous SRP, labeled by $P_2$. In this regime, the order parameters satisfy
\(\langle c_i \rangle = \langle c_j \rangle \neq \langle c_k \rangle,\,\text{for } i, j, k = 1,2,3,\)
and similarly for $\langle S_j^- \rangle$. Notably, the real parts of the order parameters for the two identical sites have opposite sign relative to the third one. Together with the $\mathbb{Z}_2$ symmetry, each of the six possible configurations is stable, the system will randomly relax into one of these symmetry-broken stationary states, depending on the initial state.

Owing to the inclusion of counter-rotating terms in the atom-photon coupling, which induce effective nonlinearities, a bistable regime appears in the strong coupling region. This is denoted as phase $D$ in the phase diagram, where two coexisting steady states correspond to the $P_1$ (homogeneous) and $P_2$ (inhomogeneous) configurations, respectively. That is, the system will evolve into one of the steady states, depending on the initial state.  Combined with the $B$ phase, we find that a homogeneous steady state described by Eq.~(\ref{hom}) can only exist in the parameter regime of $g > g_c^{P, {HSRP}}$, which spans both phases $B$ and $D$. The corresponding critical coupling strength is given by~\cite{SM}
\begin{equation}
g_c^{P,{HSRP}} = \max_k  \frac{1}{2} \left[ \frac{ \omega_a^2 \omega_{P,k} ( \kappa^2 + \omega_{P,1}^2 )^3 }{ \omega_{P,1}^3 ( \kappa^2 + \omega_{P,k}^2 ) } \right]^{1/4},
\end{equation}
and is indicated by the black dashed line in Fig.~\ref{fig:xiangtuPBC}(a).

For small values of the inter-site photon hopping strength $\xi$, the system undergoes a phase transition from the NP ($A$) to the bistable SRP ($D$), passing through the monostable homogeneous SRP ($B$), as illustrated in Fig.~\ref{fig:xiangtuPBC}(c), where the order parameters $\langle c_j \rangle$ are plotted as functions of the coupling strength $g$. This result reveals that the transition from phase $A$ to phase $B$ is a second-order phase transition, whereas the transition from phase $B$ to phase $D$ may be either a smooth crossover or a first-order phase transition, depending on which steady state is realized within the bistable regime of phase $D$. In contrast, for larger values of $\xi$, the system transitions directly from the NP $A$ to the inhomogeneous SRP $C$ via a second-order phase transition, as shown in Fig.~\ref{fig:xiangtuPBC}(d).

Under OBC, the system exhibits a significantly richer and more intricate phase diagram, as shown in Fig.~\ref{fig:xiangtuOBC}(a). In addition to the monostable SRPs ($B$, $C$, $D$), we also identify multiple bistable phases ($E$, $F$, $G$, $H$) and even a {tristable phase} ($I$). The detailed characteristics of the steady states corresponding to different phase regions are illustrated in Fig.~\ref{fig:xiangtuOBC}(b), with representative order parameter configurations denoted by $O_i$ ($i=1,2,3,4$). ($O_1$): $\langle c_1 \rangle = \langle c_3 \rangle \neq \langle c_2 \rangle \neq 0$, with the real parts of all $\langle c_j \rangle$ having the same sign. ($O_2$): $\langle c_1 \rangle = -\langle c_3 \rangle$, $\langle c_2 \rangle = 0$, indicating that the middle site remains in the NP, while the side sites exhibit superradiance. ($O_3$): $\langle c_1 \rangle \neq \langle c_2 \rangle \neq \langle c_3 \rangle \neq 0$, with the real part of $\langle c_2 \rangle$ sharing the same sign with only one of the two side sites. ($O_4$): $\langle c_1 \rangle = \langle c_3 \rangle \neq \langle c_2 \rangle \neq 0$, where the side sites share the same sign in their real parts, which is opposite to that of the middle site.

From a symmetry perspective, the steady states $O_1$, $O_3$, and $O_4$ break the global $\mathbb{Z}_2$ symmetry of the system. In $O_2$,  the middle site remains in a normal state preserving the $\mathbb{Z}_2$ symmetry locally.

To further illustrate the rich phase structure, we present two representative transition trajectories in Figs.~\ref{fig:xiangtuOBC}(c) and \ref{fig:xiangtuOBC}(d) for $\xi = 0.2\omega$ and $\xi = 0.67\omega$, respectively. Specifically, the transition sequence $A \rightarrow B \rightarrow F \rightarrow I$ and $A \rightarrow C \rightarrow D \rightarrow H$ is observed. These results indicate that while the transition from the NP to a SRP remains of second order, the transitions among different SRPs can be either smooth crossovers or first-order phase transitions, depending on the involved steady-state branches.

\begin{figure}
    \centering
    \includegraphics[width=\linewidth]{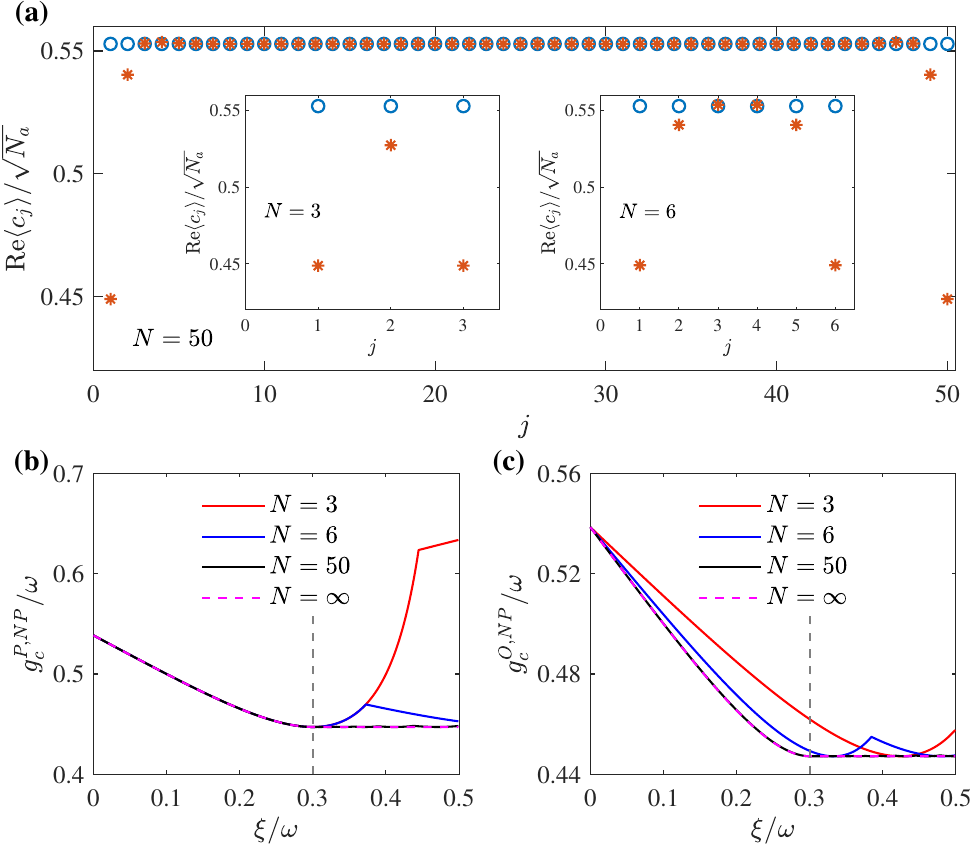}
    \caption{(a) The steady-state order parameter \( \langle c_j \rangle / \sqrt{N_a} \) for a Dicke lattice with \( N = 50 \) with PBC (empty circles) and OBC (stars). Insets show the corresponding results for \( N = 3 \) and \( N = 6 \) with \( \xi = 0.2\omega \) and \( g = 0.6\omega \). (b,c) The critical coupling strength as a function of the photon hopping \( \xi/\omega \) for different system sizes \( N \), under (b) periodic and (c) open boundary conditions, with the vertical dashed gray line representing $\xi=(\omega_c-\kappa)/2$.  Other parameters are the same as in Fig.~\ref{fig:xiangtuPBC}.
}
    \label{fig:N}
\end{figure}

{\bf Open boundary effects for finite $N\geq3$}-In most previous studies on the Rabi and Dicke lattice model~\cite{ND2014,JCL2016,Rabi2021PRL,Rabi2022PRL,
threeD2022,twoD2024,twoD2025,threeD2025}, the OBC has received relatively little attention. However, as revealed by the phase diagrams for $N=3$, we identify three novel effects induced by OBC compared to PBC. These effects persist for larger but finite system sizes $N$.

The first and most striking effect is the absence of a homogeneous SRP under OBC. This absence persists not only for the minimal case of $N=3$, but also for larger, yet finite, Dicke lattices with $3 < N < \infty$. This can be analytically understood from the steady-state solution of Eq.~(\ref{eqs:wentai}). If
$\langle c_1 \rangle = \langle c_2 \rangle = \cdots = \langle c_N \rangle$, then Eq.~(\ref{eqs:wentai}) yields
\(
-\kappa\,\mathrm{Re}(\langle c_j \rangle) + (\omega_c - \xi)\,\mathrm{Im}(\langle c_j \rangle) = 0 \quad \text{for } j=1,N,
\) and
\(
-\kappa\,\mathrm{Re}(\langle c_j \rangle) + (\omega_c - 2\xi)\,\mathrm{Im}(\langle c_j \rangle) = 0 \quad \text{for } j=2,\ldots,N-1.
\)
Clearly, these two equations cannot be simultaneously satisfied, thus ruling out the possibility of a spatially homogeneous steady-state solution under OBC for any finite $N$.

This analytical result is confirmed numerically in Fig.~\ref{fig:N}(a), where we plot the real parts of $\langle c_j \rangle$ at $t=400/\omega$ (i.e. $t=160/\kappa$)---a time sufficiently long for the system to reach its steady state---starting from a spatially homogeneous initial condition. Under PBC, the steady state remains homogeneous across all sites, i.e., $\langle c_1 \rangle = \langle c_2 \rangle = \cdots = \langle c_N \rangle$. In contrast, for OBC, even with a relatively large system size of $N=50$, the sites near the boundaries deviate significantly from those in the bulk, breaking the homogeneity. The inset subfigures of Fig.~\ref{fig:N}(a) further illustrate this point for smaller lattices with $N=3$ and $N=6$. In both cases, the steady states remain inhomogeneous under OBC, confirming that the homogeneous SRP is completely absent for finite open systems, regardless of size.

The second key effect of the OBC is the modification of the critical on-site atom-photon coupling strength that separates the SRPs from the NP. Although the critical coupling is still determined by the same equation, given by Eq.~(\ref{phaseboundary}), for both PBC and OBC, the modification arises from the difference in the dispersion relations, $\omega_{P,k}$ and $\omega_{O,k}$. Compared to the result for PBC shown in Fig.~\ref{fig:N}(b), the effects induced by the OBC become dominant in the regime $\xi < (\omega_c - \kappa)/2$, as illustrated in Fig.~\ref{fig:N}(c). In this regime, the critical coupling strength is independent of $N$ under PBC, but exhibits a strong dependence on $N$ under OBC.
As shown analytically in Eq.~(\ref{phaseboundary}), the minimum critical coupling occurs when the photon mode frequency $\omega_{P(O),k}$ is approximately equal to the decay rate of the resonators $\kappa$. For moderate dissipation ($2\kappa < \omega_c$), this minimum is taken as the $k=1$ mode in both PBC and OBC, provided $\xi < (\omega_c - \kappa)/2$. Under these conditions, the critical coupling becomes size-independent for PBC, where $\omega_{P,1} = \omega_c - 2\xi$, while it remains size-dependent for OBC, where $\omega_{O,1} = \omega_c - 2\xi \cos[\pi/(N+1)]$. In contrast, when $\xi > (\omega_c - \kappa)/2$, the mode whose frequency best satisfies $\omega_{P(O),k} \simeq \kappa$ depends on the system size $N$, as detailed in the SM~\cite{SM}. Consequently, the critical coupling strength becomes $N$-dependent for both PBC and OBC.

We also note that the effect of the OBC on the critical coupling strength gradually diminishes as $N$ increases. As shown in Figs.~\ref{fig:N}(b) and (c), for $N=50$, the critical atom-photon coupling becomes nearly insensitive to the boundary condition and approaches the result in the limit of $N \rightarrow \infty$, which is given by~\cite{SM,ND2014}
\begin{equation}
g_c^{P(O),NP}\big|_{N=\infty} =
\begin{cases}
\frac{1}{2}\sqrt{\omega_a \omega_{P,1} \left( 1 + \frac{\kappa^2}{\omega_{P,1}^2} \right)}, & 0 < \xi < \frac{\omega_c - \kappa}{2}\\[6pt]
\sqrt{ \frac{\omega_a \kappa}{2} }, & \frac{\omega_c - \kappa}{2} < \xi < \frac{\omega_c}{2}
\end{cases}.
\end{equation}

The final notable effect concerns the stability boundary of the NP, which is determined by the range of hopping strength $\xi$ that still supports a dynamically stable NP. For the case of $N=3$, the dispersion relation provided in the SM~\cite{SM} indicates that the allowed range of $\xi$ is $-\omega_c < \xi < \omega_c/2$ under PBC and $-\omega_c/\sqrt{2} < \xi < \omega_c/\sqrt{2}$ under OBC. However, as $N$ increases, the distinction between the two boundary conditions gradually vanishes, and the permissible parameter regime asymptotically converges to the same range.

\begin{figure}
    \centering
    \includegraphics[width=\linewidth]{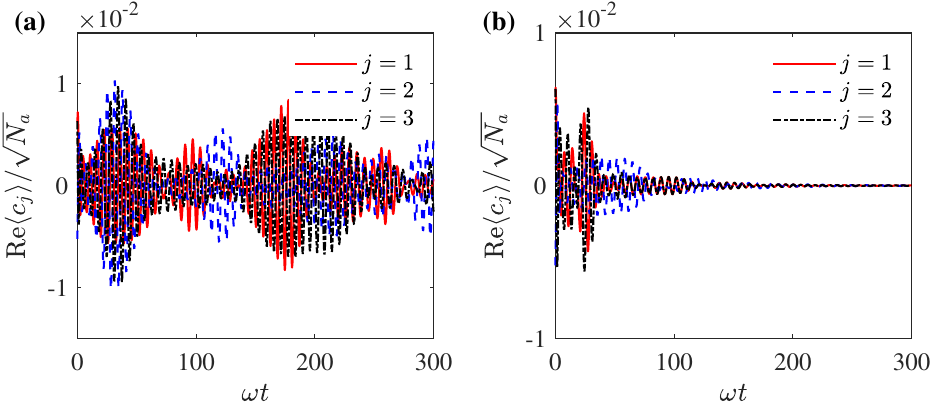}
    \caption{Time evolution of the order parameters \( \langle c_j \rangle / \sqrt{N_a} \) for (a) periodic boundary conditions (PBC) and (b) open boundary conditions (OBC). Here, \( \xi = 0.6\omega \), \( g = 0.3\omega \), and \( N = 3 \). Other parameters are the same as in Fig.~\ref{fig:xiangtuPBC}.
}
    \label{fig:NPDM}
\end{figure}

To demonstrate this effect for $N=3$, we numerically simulate the photon dynamics by plotting the time evolution of the real parts of $\langle c_j \rangle$, starting from randomly chosen initial conditions in the intermediate coupling regime  of $\omega_c/2<\xi<\omega_c/\sqrt{2}$  in Fig.~\ref{fig:NPDM}. Under PBC, the order parameters exhibit persistent oscillations and never settle into a steady state, indicating dynamical instability. In contrast, for OBC, the same initial conditions lead to transient oscillations that eventually decay, with all $\langle c_j \rangle$ approaching zero, signifying a stable NP.

{\bf Conclusion}-In summary, we have investigated how boundary conditions influence dissipative phase transitions in light-matter interacting systems, using the Dicke lattice model as a paradigmatic example. Our results reveal that OBC not only shift the critical on-site atom-photon coupling strength but also alter the range of inter-site photon hopping strengths that ensure dynamical stability in finite lattices. Notably, a fully homogeneous SRP is entirely absent under OBC, irrespective of system size.

Our predictions are particularly relevant for near-term experimental realizations of such models, which are naturally limited in size and where different phases can be explored by simply adjusting one boundary. Apart from quantum optical settings with cold atoms~\cite{Dicke20102,DickeEX2010,DickeEX2014,DickeEX2015,HRR2013} and trapped ions~\cite{KA2021}, potential platforms to study such photonic lattice models in particular include arrays of superconducting microwave resonators, coupled (ultra-)strongly to superconducting qubits~\cite{SC2010,SC2012,SC2023,ultra1,ultra6,ultra7,AV2022a},
 clouds of Rydberg atoms~\cite{SD2012,MP2014,DY2016,SG2019,AA2020,TZ2025,BW2025}, or ensembles of NV centers~\cite{ND2014,NV2022,DI2010prl,YK2010prl,RA2011prl}.

Finally, our study on one-dimensional lattice model provides a foundation for generalizing to higher-dimensional geometries such as square, cubic, or hexagonal lattices---paving the way to explore boundary-induced effects on
quantum phase transitions in both natural and engineered quantum materials with light-matter interactions.

Z. W. is supported by Science and Technology Development Project of Jilin Province (Grant No. 20230101357JC) and  National Natural Science Foundation of China (Grant No. 12375010). Q. H. is supported by Beijing Natural Science Foundation (Grant No. Z240007) and National Natural Science Foundation of China (Grant No. 12125402). P. R. acknowledges support from the Deutsche Forschungsgemeinschaft (DFG, German Research Foundation)-522216022. This research is part of the Munich Quantum Valley, which is supported by the Bavarian state government with funds from the Hightech Agenda Bayern Plus.


\end{document}